\documentclass[fdp,a4paper,fleqn%
]{w-art}
\usepackage{times,cite,w-thm}
\theoremstyle{plain}

\theoremstyle{definition}

\usepackage[]{graphicx}

\newcommand{\scriptmath}[1]{{\scriptsize{\mbox{$#1$}}}}

\usepackage{amsmath,amssymb,amsfonts,amsthm,bbm,cancel,wasysym,array,booktabs,rotating}

\newcommand{\units}[1]{~\mathrm{#1}}

\begin{document}
\DOIsuffix{theDOIsuffix}
\Volume{55}
\Month{01}
\Year{2007}
\pagespan{1}{}
\Receiveddate{XXXX}
\Reviseddate{XXXX}
\Accepteddate{XXXX}
\Dateposted{XXXX}
%
%



\title[Electroweak constraints]{Electroweak constraints on new physics}


\author[F. del Aguila]{F. del Aguila $^a$%
  \footnote{E-mail:~\textsf{faguila@ugr.es} 
}}
\address[]{$^a$ CAFPE and Depto. de F\'{\i}sica Te\'orica y del 
Cosmos, Universidad de Granada, E-18071 Granada, Spain \\ 
$^b$ Department of Physics, University of Notre Dame, 
Notre Dame, Indiana 46556, USA}
\author[J. de Blas]{J. de Blas $^b$%
  \footnote{E-mail:~\textsf{jdeblasm@nd.edu}}}
\begin{abstract}
We briefly review the limits on new interactions implied by 
electroweak precision data. Special attention is payed to 
the bounds on the Higgs boson mass. 
We also comment on the required cancellation among the 
new contributions to precisely measured electroweak observables 
in any Standard Model extension, if the new particles have to 
evade the indirect constraints on their couplings and masses 
but still remain at the LHC reach.  
\end{abstract}
\maketitle                   





\section{Introduction}
\label{Introduction}

The time for establishing the mechanism of the Standard Model 
(SM) symmetry breaking seems to have arrived with the 
large hadron collider (LHC) era. 
Thus, it is widely believed that the SM is a low energy effective theory and 
that new physics must exist near the TeV scale which makes natural the 
observed values of the gauge boson and fermion masses. 
However, the four scenarios with or without the SM Higgs and/or 
new physics observed at LHC are still possible.  
Although it shall be paradoxical that electroweak precision 
data (EWPD) are in agreement with the SM predictions at the few per 
mille level 
\cite{Nakamura:2010zzi,:2010vi}, 
implying that the new physics scale is relatively large, 
but new resonances other than the SM Higgs are detected at the LHC 
\cite{Barbieri:2000gf}. 
EWPD also disfavor a SM Higgs mass much larger than its 
present direct limit and that no new physics is found up to a few TeV 
\cite{deBlasThesis} (see below).  
At any rate, EWPD and direct searches are complementary and 
whatever physics LHC reveals, it shall fulfill the indirect constraints. 

Physics could be unexpected but we shall assume that we will 
be finally left with the SM plus some new particles with masses 
above the electroweak scale. Such a scenario can be described 
for energies below a few hundreds of GeV by an effective Lagrangian 
with the SM fields and gauge symmetry. The new physics being 
encoded in the operators of dimension $d > 4$. In the following we 
update the limits on these operators, assuming universality and 
taking one at a time. In general, only those contributing to observables 
showing (small) deviations from the SM predictions are not 
suppressed at the per cent level. We then explain how to accommodate 
a large Higgs mass, which is the only SM parameter still unknown. 
What can be done invoking new heavy neutrinos and/or vector bosons.   
Finally, we comment on how to fulfill the EWPD constraints and 
still allow for new resonances at the LHC reach.   


\section{Effective Lagrangian approach, EWPD and model independent limits}
\label{Limits}

Let us write the effective Lagrangian with the SM fields and symmetries 
\begin{equation}
{\cal L}_{\mathrm{eff}}=\sum_{d=4}^\infty \frac{1}{\Lambda^{d-4}}{\cal L}_d=
{\cal L}_4+\frac {1}{\Lambda}{\cal L}_5+\frac {1}{\Lambda^2}{\cal L}_6+\dots, \quad 
{\cal L}_d=\sum_i {\alpha_i^d} {\cal O}_i^d, \quad \left[{\cal O}_i^d\right]=d,
\label{GenEffLag}
\end{equation}
where $\Lambda$ is the (unknown) cutoff scale up to which the effective 
Lagrangian description is valid, and each ${\cal L}_d$ contains all the local 
operators of canonical mass dimension $d$ allowed by the symmetries. 
(${\cal L}_5$ only contains one operator 
\cite{Weinberg:1979sa,Buchmuller:1985jz}, 
which violates lepton number and can be neglected because it is proportional 
to the very tiny neutrino masses and then plays no r\^ole in our analysis 
\cite{Mohapatra:2005wg}.) 
The operators of dimension six ${\cal O}_i$ are classified in 
\cite{Buchmuller:1985jz}
(see for a non-redundant set 
\cite{Grzadkowski:2003tf}). 
In the table 
we update the limits on their coefficients 
$\alpha_i/\Lambda^2$. 
The data included in the fit are described in 
\cite{deBlasThesis,delAguila:2008pw,delAguila:2009vv,delAguila:2010mx,delAguila:2011yd}, 
but updated to their more recent values. 
We separate them in four sets and collect the corresponding bounds in the 
first four columns. 
The global fit to all data is gathered in the last column. 
We assume universality and the fits are performed adding one operator at a 
time to the SM.   

\begin{table}[]
\begin{center}
\begin{tabular}{ l c c  c  c  c  c }
\hline
\multicolumn{2}{c}{Operator}&$Z$ pole& $W$ data&Low Energy&LEP 2&Global fit\\
\multicolumn{2}{c}{coefficient\!}\!&\multicolumn{5}{c}{$95\%$ C.L. limits [TeV$^{-2}$]}\\
\hline
&&&&&&\\[-0.375cm]
\!\!LLLL
&$\!\!\!\!\!\frac{\alpha_{ll}^{(1)}}{\Lambda^2}$\!&-&- 
&$\!\!\left[-0.034,0.245\right]\!\!$ &$\!\!\left[-0.089,0.025\right]\!\!$ &$\!\!\left[-0.065,0.040\right]\!\!$ \\
&&&&&&\\[-0.4cm]
&$\!\!\!\!\!\frac{\alpha_{ll}^{(3)}}{\Lambda^2}$\!&$\!\!\left[-0.007,0.056\right]\!\!$
&$\!\!\left[-0.007,0.009\right]\!\!$ &$\!\!\!\!\left[-0.066,0.047\right]\!\!\!\!$ &$\!\!\left[-0.045,0.042\right]\!\!$ 
&$\!\!\left[-0.006,0.006\right]\!\!$ \\
&&&&&&\\[-0.4cm]
&$\!\!\!\!\!\frac{\alpha_{lq}^{(1)}}{\Lambda^2}$\!&-&- &$\!\!\left[-0.028,0.019\right]\!\!$ 
&$\!\!\left[\phantom{+}0.011,0.437\right]\!\!$ &$\!\!\left[-0.025,0.021\right]\!\!$ \\
&&&&&&\\[-0.4cm]
&$\!\!\!\!\!\frac{\alpha_{lq}^{(3)}}{\Lambda^2}$\!&-&$\!\!\left[-0.007,0.009\right]\!\!$ 
&$\!\!\left[-0.023,0.070\right]\!\!$ &$\!\!\left[\phantom{+}0.001,0.057\right]\!\!$ &$\!\!\left[-0.004,0.011\right]\!\!$ \\
&&&&&&\\[-0.375cm]
\hline
&&&&&&\\[-0.375cm]
\!\!RRRR
&$\!\!\!\!\!\frac{\alpha_{ee}}{\Lambda^2}$\!&-&- 
&$\!\!\left[-0.257,0.031\right]\!\!$ &$\!\!\left[-0.056,0.012\right]\!\!$ &$\!\!\left[-0.060,0.006\right]\!\!$ \\
&&&&&&\\[-0.4cm]
&$\!\!\!\!\!\frac{\alpha_{eu}}{\Lambda^2}$\!&-&- &$\!\!\left[-0.037,0.061\right]\!\!$ 
&$\!\!\left[~\!\!-0.196~\!\!,~\!\!-0.003\right]\!\!$ &$\!\!\left[-0.055,0.033\right]\!\!$ \\
&&&&&&\\[-0.4cm]
&$\!\!\!\!\!\frac{\alpha_{ed}}{\Lambda^2}$\!&-&- &$\!\!\left[-0.038,0.051\right]\!\!$ 
&$\!\!\left[\phantom{+}0.004,0.262\right]\!\!$ &$\!\!\left[-0.022,0.062\right]\!\!$ \\
&&&&&&\\[-0.375cm]
\hline
&&&&&&\\[-0.375cm]
\!\!LRRL
&$\!\!\!\!\!\frac{\alpha_{le}}{\Lambda^2}$\!&-&- 
&$\!\!\left[-1.146,1.132\right]\!\!$ &$\!\!\left[-0.059,0.101\right]\!\!$ &$\!\!\left[-0.059,0.101\right]\!\!$ \\
&&&&&&\\[-0.4cm]
&$\!\!\!\!\!\frac{\alpha_{lu}}{\Lambda^2}$\!&-&- &$\!\!\left[-0.082,0.108\right]\!\!$ 
&$\!\!\left[\phantom{-}0.003,0.955\right]\!\!$ &$\!\!\left[-0.062,0.123\right]\!\!$ \\
&&&&&&\\[-0.4cm]
&$\!\!\!\!\!\frac{\alpha_{ld}}{\Lambda^2}$\!&-&- &$\!\!\left[-0.072,0.104\right]\!\!$ 
&$\!\!\left[~\!\!-1.273~\!\!,~\!\!-0.003\right]\!\!$ &$\!\!\left[-0.084,0.091\right]\!\!$ \\
&&&&&&\\[-0.4cm]
&$\!\!\!\!\!\frac{\alpha_{qe}}{\Lambda^2}$\!&-&- &$\!\!\left[-0.056,0.038\right]\!\!$ 
&$\!\!\left[\phantom{+}0.011,0.566\right]\!\!$ &$\!\!\left[-0.047,0.045\right]\!\!$ \\
&&&&&&\\[-0.375cm]
\hline
&&&&&&\\[-0.375cm]
\!\!SVF
&$\!\!\!\!\!\frac{\alpha_{\phi l}^{(1)}}{\Lambda^2}$\!
&$\!\!\left[-0.004,0.009\right]\!\!$&- &$\!\!\left[-0.078,0.027\right]\!\!$ 
&$\!\!\left[-0.036,0.172\right]\!\!$ &$\!\!\left[-0.004,0.009\right]\!\!$ \\
&&&&&&\\[-0.4cm]
&$\!\!\!\!\!\frac{\alpha_{\phi q}^{(1)}}{\Lambda^2}$\!&$\!\!\left[-0.021,0.033\right]\!\!$&- 
&$\!\!\left[-0.019,0.028\right]\!\!$ &$\!\!\left[-0.004,1.160\right]\!\!$ &$\!\!\left[-0.012,0.023\right]\!\!$ \\
&&&&&&\\[-0.4cm]
&$\!\!\!\!\!\frac{\alpha_{\phi e}^{(1)}}{\Lambda^2}$\!&$\!\!\left[-0.011,0.006\right]\!\!$&- 
&$\!\!\left[-0.152,0.047\right]\!\!$ &$\!\!\left[-0.235,0.102\right]\!\!$ &$\!\!\left[-0.012,0.005\right]\!\!$ \\
&&&&&&\\[-0.4cm]
&$\!\!\!\!\!\frac{\alpha_{\phi u}^{(1)}}{\Lambda^2}$\!&$\!\!\left[-0.053,0.067\right]\!\!$&- 
&$\!\!\left[-0.032,0.063\right]\!\!$ &$\!\!\left[-0.049,2.821\right]\!\!$ &$\!\!\left[-0.024,0.050\right]\!\!$ \\
&&&&&&\\[-0.4cm]
&$\!\!\!\!\!\frac{\alpha_{\phi d}^{(1)}}{\Lambda^2}$\!&$\!\!\left[-0.131,0.031\right]\!\!$&- 
&$\!\!\left[-0.039,0.049\right]\!\!$ &$\!\!\left[-3.762,0.065\right]\!\!$ &$\!\!\left[-0.047,0.031\right]\!\!$ \\
&&&&&&\\[-0.4cm]
&$\!\!\!\!\!\frac{\alpha_{\phi l}^{(3)}}{\Lambda^2}$\!&$\!\!\left[-0.011,0.006\right]\!\!$
&$\!\!\left[-0.009,0.005\right]\!\!$ &$\!\!\left[-0.023,0.056\right]\!\!$ &$\!\!\left[-0.178,0.005\right]\!\!$ 
&$\!\!\left[-0.006,0.004\right]\!\!$ \\
&&&&&&\\[-0.4cm]
&$\!\!\!\!\!\frac{\alpha_{\phi q}^{(3)}}{\Lambda^2}$\!&$\!\!\left[-0.008,0.011\right]\!\!$
&$\!\!\left[-0.009,0.007\right]\!\!$ &$\!\!\left[-0.070,0.023\right]\!\!$ &$\!\!\left[\!\!\phantom{-}3\!\cdot\!\!10^{-4}\!,0.308\right]\!\!$ 
&$\!\!\left[-0.006,0.006\right]\!\!$ \\
&&&&&&\\[-0.4cm]
&$\!\!\!\!\!\frac{\alpha_{\phi ud}}{\Lambda^2}$\!&-&$\!\!\left[-0.015,0.018\right]\!\!$ 
&$\!\!\left[-0.071,0.156\right]\!\!$ &-&$\!\!\left[-0.014,0.019\right]\!\!$ \\
&&&&&&\\[-0.375cm]
\hline
&&&&&&\\[-0.375cm]
\!\!Oblique
&$\!\!\!\!\!\frac{\alpha_{\phi}^{(3)}}{\Lambda^2}$\!
&$\!\!\left[-0.112,0.013\right]\!\!$&$\!\!\left[~\!\!-0.120~\!\!,~\!\!-0.002\right]\!\!$ &$\!\!\left[-0.069,0.139\right]\!\!$ 
&$\!\!\left[-0.406,0.026\right]\!\!$&$\!\!\left[-0.107,0.001\right]\!\!$ \\
&&&&&&\\[-0.4cm]
&$\!\!\!\!\!\frac{\alpha_{W\!B}}{\Lambda^2}$\!&$\!\!\left[-0.017,0.005\right]\!\!$
&$\!\!\!\left[~\!\!-0.056,~\!\!-7\!\cdot\!\!10^{-4}\right]\!\!\!$ &$\!\!\left[-0.010,0.061\right]\!\!$ 
&$\!\!\left[-0.181,0.014\right]\!\!$ &$\!\!\left[-0.010,0.003\right]\!\!$ \\
&&&&&&\\[-0.375cm]
\hline
\end{tabular}
\caption{$95\%$ C.L. limits on ($90\%$ confidence interval of) the 
dimension six operator coefficients entering in EWPD. The limits are obtained 
from a fit considering only one operator at a time and for each data set. 
Limits are in units of TeV$^{-2}$. The different columns show the results for 
different fits depending on the observables included. The second column 
($W$ data) also includes the constraints from CKM universality. When a given operator contributes
to a physical process from which any of the SM inputs is derived, it indirectly corrects the predictions for all electroweak observables (e.g., the operator
${\cal O}_{ll}^{(3)}$ which modifies 
the prediction for the muon decay constant $G_\mu$).}
\label{table: indivOfit}
\end{center}
\end{table}

As can be observed, some of the most significant departures ($\sim1~\sigma$ or larger) 
from the SM predictions can be eased with few of these operators. This translates into asymmetric intervals. 
For instance, 
the excess of the hadronic cross section observed at LEP 2 can be explained 
by four-fermion operators involving electrons and quarks, 
like ${\cal O}^{(1,3)}_{lq}$, etcetera. 
Parity violation in M{\o}ller scattering can be improved by ${\cal O}^{(1)}_{ll}$ 
or ${\cal O}_{ee}$, for example. 
On the other hand, the relatively large value of the $W$ mass can be accounted 
by ${\cal O}^{(3)}_{\phi}$ and ${\cal O}_{W B}$. 
While the large forward-backward bottom asymmetry results in 
an asymmetric ${\alpha}^{(1)}_{\phi d}$ 
interval, although we assume universality. 
At any rate, the size and asymmetry of the intervals get reduced when all data 
are considered.


\section{Implications on the Higgs mass}
\label{Higgs}

In the previous fits to dimension six operators the SM parameters are fixed 
to their best value in the fit to the SM alone, except for the Higgs mass 
$M_H$ which is left free. 
This, in general, prefers to be next to its direct lower limit of 114 GeV 
\cite{Nakamura:2010zzi,:2010vi,Barate:2003sz}.
In the left figure we show the $\chi^2$ dependence on $M_H$ in the fit 
to the SM alone with all SM parameters free (upper black solid line). 
As it is apparent, if the SM Higgs is found to be relatively heavy, 
further physics has to cancel its one-loop contributions to the different 
electroweak precision observables, in order
to restore the excellent agreement with the data. 
In particular, it has to balance the negative quantum correction to 
the $\rho = M_W^2 / M_Z^2 \cos^2{\theta_W}$ parameter \cite{Peskin:2001rw}.  
This can be done at tree level increasing the numerator or decreasing the 
denominator, yielding in both cases the required positive contribution. 
The former can be effectively achieved reducing the SM contribution to the Fermi 
constant $G_\mu$ by mixing the electron neutrino with a sterile heavy 
neutrino $N_e$ 
\cite{delAguila:2008pw,delAguila:2009bb}, 
and the latter mixing the $Z^0$ boson with heavier extra vector bosons 
\cite{Chanowitz:2008ix,delAguila:2010mx}. 
\footnote{
Note that in the operator basis chosen here
corrections to $G_\mu$ can be encoded either 
in ${\cal O}_{\phi l}^{(3)}$ or ${\cal O}_{ll}^{(1,3)}$. While, 
direct corrections to the $\rho$ parameter are accounted 
by ${\cal O}_{\phi}^{(3)}$, allowing for large $M_H$ values for 
negative $\alpha_\phi^{(3)}$ (see Table \ref{table: indivOfit}).}
In the left figure we show the effect of both possibilities.  
The second upper line (blue dashed) corresponds to the heavy neutrino 
addition, which can not completely account for a heavy Higgs 
but improves the global fit. 
Whereas there are two gauge boson additions balancing the heavy 
Higgs corrections to EWPD, named ${\cal B}$ and ${\cal W}^1$ in 
\cite{delAguila:2010mx}, 
respectively (bottom green dotted-dashed and second bottom red 
solid lines in the figure).  
In these three fits the only SM parameters left free, besides the 
Higgs mass, are the strong coupling constant and the top mass. 
The large $\chi^2$ values on the ordinate reminds the large number (212) of 
data included in the fits. 
Finally, in the right figure we plot the same as in the left one but 
replacing the present large collider bounds 
\cite{Barate:2003sz}
by two guesses of the Higgs mass eventually measured at CERN 
\cite{Vickey:2008ww}. 

\begin{figure}
\input{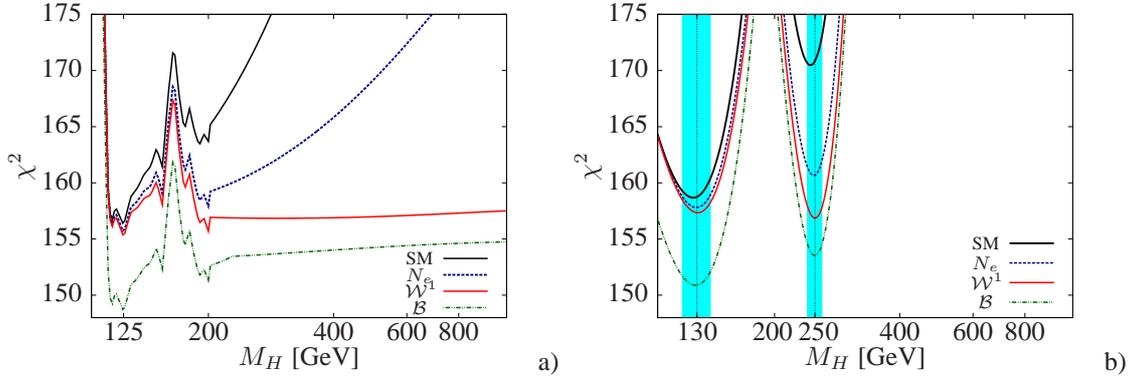}
\caption{\textbf{a)} From top to bottom, minimum $\chi^2$ as a function 
of the Higgs mass for the SM fit, and the fits including besides a heavy 
neutrino singlet $N_e$ coupled to the first lepton family, a vector triplet 
of hyperchage one ${\cal W}^1$, and a neutral vector singlet ${\cal B}$ 
(see \cite{deBlasThesis,delAguila:2008pw,delAguila:2010mx} for conventions).
\textbf{b)} The same but assuming that the SM Higgs is found to have a mass 
$M_H=130\pm10\units{GeV}$ or $M_H=250\pm10\units{GeV}$ (blue bands).}
\label{fig:5}
\end{figure}


\section{Cancelling contributions from extended spectra}
\label{Cancelling}

The previous fits make apparent the paradox that the SM describes 
physics up to the LEP 2 energy ($\sim$ 209 GeV) with a precision in general below the 
per cent level, but we still expect that LHC will discover further resonances 
near the TeV 
\cite{Barbieri:2000gf}.   
If so, a model dependent pattern of cancellations must arise resulting 
in small contributions to electroweak precision observables. 
The corresponding discussion for the case of extra gauge bosons is 
presented in \cite{delAguila:2010mx,delAguila:2011yd}. 
Examples with cancellations based on custodial symmetries can be 
found in \cite{Agashe:2006at} for extra quarks or in 
\cite{Agashe:2009tu} 
for extra leptons. 
However, in these models flavor plays an essential 
r\^ole because the new fermions mainly mix with the third family, 
as may be in Nature.

\begin{acknowledgement}
We are grateful to the Corfu Institute 2010 organizers for their habitual 
kind hospitality, 
and to M. P\'erez-Victoria for collaboration in the work reviewed here. 
This work has been partially supported by MICINN  (FPA2006-05294 and 
FPA2010-17915) and by Junta de Andaluc\'{\i}a (FQM 101, FQM 3048 and 
FQM 6552). 
The work of J.B. has been supported in part by the U.S. National Science 
Foundation under Grant PHY-0905283-ARRA. 
\end{acknowledgement}


%
%

\end{document}